\documentclass[pra,aps,twocolumn,superscriptaddress,showpacs]{revtex4}

\usepackage{psfig}

\begin{document}

\title{Optimal probabilistic cloning and purification  of quantum states}

\author{Jarom\'{\i}r Fiur\'{a}\v{s}ek}
\affiliation{QUIC, Ecole Polytechnique, CP 165, 
Universit\'{e} Libre de Bruxelles, 1050 Bruxelles, Belgium }
\affiliation{Department of Optics, Palack\'{y} University, 
17. listopadu 50, 77200 Olomouc, Czech Republic}

\begin{abstract}
We investigate the probabilistic cloning and purification of quantum
states. The performance of these probabilistic operations is quantified
by the average fidelity between the ideal and actual output states.
We provide a simple formula for the 
maximal achievable average fidelity and we explictly show how to construct 
a probabilistic operation that achieves
this fidelity. We illustrate our method on several examples such as 
the phase covariant cloning of qubits, cloning of coherent states, and
purification of qubits transmitted via depolarizing channel and amplitude
damping channel. Our examples reveal that the probabilistic cloner may
yield higher fidelity than the best deterministic cloner even when the states
that should be cloned are linearly dependent and are drawn from 
a continuous set.
\end{abstract}

\maketitle

\section{Introduction}

The recent spectacular development of the quantum information theory
has revealed that information processing based on the laws of quantum
mechanics enables implementation of tasks  that are impossible or very hard 
to accomplish classically. The prime examples are
the unconditionally secure cryptography \cite{Gisin02} 
and the exponential speedup of certain computational tasks, such as the 
factoring of integers \cite{Shor94}. 
On the other hand, the linearity of quantum mechanics also 
imposes certain constraints on the processing of quantum information 
that have no classical counterpart.
Perhaps the most famous example is the no-cloning theorem which states
that an unknown quantum state cannot be copied \cite{Wootters82}.  However, 
this restriction provides, in fact, a valuable resource
explored in the quantum key distribution
protocols, because it forbids an eavesdropper Eve to gain information
on the distributed secret key without introducing errors.

Since exact copying is forbidden, a natural problem arises
what is the optimal approximate cloning transformation.
This question was first asked by Bu\v{z}ek and Hillery in their seminal paper
\cite{Buzek96} and since then it has been addressed by numerous authors 
who considered various cloning scenarios, such as cloning of qubits
\cite{Gisin97,Bruss98}, 
cloning of d-dimensional systems (qudits) 
\cite{Werner98,Buzek98,Keyl99,Cerf01,Braunstein01}
and cloning of continuous variables
\cite{Cerf00,Braunstein01CV,Fiurasek01CV}.
Much attention has been recently paid also to the cloning of
subsets of Hilbert space, such as the phase
covariant cloning machine for equatorial qubits
\cite{Bruss00,DAriano01,DAriano03,Fiurasek03} 
and cloning of real states \cite{Navez03} or maximally entangled states 
\cite{Lamoureux03}.
Remarkably, the cloning machines turned out to be very efficient 
or even optimal eavesdropping attacks on many quantum cryptographic 
protocols \cite{Fuchs97,Niu99,Cerf02,Acin03} and their 
investigation is largely motivated 
by these practical aspects.

Typically, the cloner is assumed to be a deterministic machine that
always produces an output. Nevertheless, one can consider also probabilistic
cloning machines that sometimes fail and do not provide any outcome.
The probabilistic cloners have been discussed in the
literature in the context of cloning of a discrete finite set of quantum
states and it was shown that an exact probabilistic cloning
is possible if and only if the set consists of linearly independent
states \cite{Duan98,Chefles98}.

However, one may hope that the probabilistic machines
may yield better results also for the sets of linearly dependent
quantum states and even for infinite (continuous) sets. 
Here, we  investigate the probabilistic cloning of linearly dependent states
and we establish a general theory of the optimal probabilistic cloning machines. 
We provide a simple formula for the optimal average fidelity of the probabilistic 
machine and we also show  how to construct the optimal 
cloning transformations.

In fact, our formalism is very general
and it concerns optimal probabilistic implementations of arbitrary
transformations whose outputs should ideally be pure states. Besides cloning,
this includes also universal NOT gate for qubits \cite{Gisin99,Buzek99NOT}, 
and, perhaps even more importantly, probabilistic purification of mixed 
quantum states \cite{Cirac99,Keyl01,Ricci04}.
In what follows we first establish the general formalism and then we
work out several explicit examples that will illustrate our method.

\section{Optimal probabilistic transformations}

The probabilistic machines investigated in the present paper optimally
(in a sense defined below) approximate the map from a set $S_{\mathrm{in}}$
of input (generally mixed) states $\rho_{\mathrm{in}}$  to the set 
$S_{\mathrm{out}}$
of the output pure states $|\psi_{\mathrm{out}}\rangle$,
\begin{equation}
\rho_{\mathrm{in}} \rightarrow \psi_{\mathrm{out}}(\rho_{\mathrm{in}}),
\label{inout}
\end{equation}
where $\psi_{\mathrm{out}}\equiv |\psi_{\mathrm{out}}\rangle\langle
 \psi_{\mathrm{out}}|$
is a short hand notation for the density matrix  of a pure state.
If $S_{\mathrm{in}}$  is a set of pure states
(such as in the case of cloning) then we replace $\rho_{\mathrm{in}}$ with
$\psi_{\mathrm{in}}$. The most general probabilistic transformation in quantum
mechanics is a linear trace decreasing completely positive (CP) map
\cite{Assumption} that transforms operators on the input Hilbert space 
${\mathcal{H}}_{\mathrm{in}}$ onto the operators on the output Hilbert space 
${\mathcal{H}}_{\mathrm{out}}$.  
According to the Jamiolkowski isomorphism \cite{Jamiolkowski72},  
any CP map $\rho_{\mathrm{out}}=\mathcal{E}(\rho_{\mathrm{in}})$
can be represented by a positive semidefinite operator $E$
on the Hilbert space 
${\mathcal{H}}={\mathcal{H}}_{\mathrm{in}}\otimes {\mathcal{H}}_{\mathrm{out}}$.
Let  $|j\rangle$ denote a basis in a $d$-dimensional Hilbert space 
${\mathcal{H}}_{\mathrm{in}}$. The operator $E$ can be obtained from the 
maximally entangled  state on 
${\mathcal{H}}_{\mathrm{in}}\otimes {\mathcal{H}}_{\mathrm{in}}$, 
$|\Phi\rangle=\sum_{j=1}^d|j\rangle_A|j\rangle_B$, if we apply the map
$\mathcal{E}$ to one part of $|\Phi\rangle$. We have
$E={\mathcal{I}}_A \otimes {\mathcal{E}}_B (\Phi)$ where $\mathcal{I}$ 
denotes the identity map.

The transformation $\rho_{\mathrm{out}}={\mathcal{E}}(\rho_{\mathrm{in}})$ 
can be rewritten  in terms of $E$ as follows,
\begin{equation}
\rho_{\mathrm{out}}= {\mathrm{Tr}}_{\mathrm{in}} [E \rho_{\mathrm{in}}^T 
\otimes \openone_{\mathrm{out}}],
\label{CPmap}
\end{equation}
where $\mathrm{Tr}_{\mathrm{in}}$ denotes the partial trace over the input
Hilbert space and $T$ stands for the transposition in the basis $|j\rangle$.
The map must be trace decreasing which means that
${\mathrm{Tr}}[\rho_{\mathrm{out}}]\leq {\mathrm{Tr}}(\rho_{\mathrm{in}})$ 
for all $\rho_{\mathrm{in}}\geq 0$. This implies that $E$  must satisfy the 
inequality
\begin{equation}
{\mathrm{Tr}}_{\mathrm{out}} (E)\leq \openone_{\mathrm{in}},
\label{tracedecreasing}
\end{equation}
where $\openone$ denotes the identity operator and the equality in
(\ref{tracedecreasing}) is achieved by deterministic (trace preserving) CP maps.

Consider a particular input $\rho_{\mathrm{in}}$. The normalized output
state $\tilde{\rho}_{\mathrm{out}}$  is given by
$\tilde{\rho}_{\mathrm{out}}=\rho_{\mathrm{out}}/P(\rho_{\mathrm{in}})$ where
\begin{equation}
P(\rho_{\mathrm{in}})={\mathrm{Tr}}[E \rho_{\mathrm{in}}^T \otimes 
\openone_{\mathrm{out}}]
\label{Prhoin}
\end{equation}
is the probability of successful application of the map $\mathcal{E}$ 
to $\rho_{\mathrm{in}}$.
The performance of the map $\mathcal{E}$ for the particular input 
$\rho_{\mathrm{in}}$ can be conveniently quantified by the fidelity 
between the actual and the ideal outputs:
\begin{equation}
F(\rho_{\mathrm{in}})=\langle \psi_{\mathrm{out}} |
\tilde{\rho}_{\mathrm{out}}|\psi_{\mathrm{out}}\rangle.
\label{Frhoindef}
\end{equation}
Expressed in terms of $E$ we have
\begin{equation}
F(\rho_{\mathrm{in}})= \frac{1}{P(\rho_{\mathrm{in}})} 
{\mathrm{Tr}}(E \, \rho_{\mathrm{in}}^T\otimes \psi_{\mathrm{out}}).
\label{Frhoin}
\end{equation}

We assume that the set $S_{\mathrm{in}}$ is endowed with an a-priori 
probability distribution $d \rho_{\mathrm{in}}$ such that 
$\int_{S_{\mathrm{in}}} d\rho_{\mathrm{in}}=1$.
Here and in what follows we assume
that the set $S_{\mathrm{in}}$ is continuous. Of course, all formulas remain 
valid also for discrete sets, one simply has to replace the integrals with
corresponding summations over the elements of $S_{\mathrm{in}}$.

The average probability of success is defined as
\begin{equation}
\bar{P}=\int_{S_{\mathrm{in}}} P(\rho_{\mathrm{in}}) d\rho_{\mathrm{in}}
=\mathrm{Tr}[E A],
\label{P}
\end{equation}
where 
\begin{equation}
A=\int_{S_{\mathrm{in}}} \rho_{\mathrm{in}}^T\otimes \openone_{\mathrm{out}}
 \,  d\rho_{\mathrm{in}}.
 \label{lambda}
\end{equation}
We now introduce the mean fidelity $F$ of the transformation $\mathcal{E}$
as the average of the fidelities $F(\rho_{\mathrm{in}})$, with proper weights
$P(\rho_{\mathrm{in}})d\rho_{\mathrm{in}}/\bar{P}$,
\begin{equation}
F=\int_{S_{\mathrm{in}}} F(\rho_{\mathrm{in}}) 
\frac{P(\rho_{\mathrm{in}})}{\bar{P}} \, d\rho_{\mathrm{in}}.
\label{F}
\end{equation}
The mean fidelity is the figure of merit considered in the present paper
and in what follows we shall look for the optimal map $\mathcal{E}$ that
maximizes $F$.

If we insert the expression (\ref{Frhoin}) into Eq. 
(\ref{F})  we obtain $F=\bar{F}/\bar{P}$ where $\bar{F}={\mathrm{Tr}}[ER]$ and
\begin{equation}
R=\int_{S_{\mathrm{in}}} \rho_{\mathrm{in}}^T \otimes \psi_{\mathrm{out}} 
d\rho_{\mathrm{in}}.
\label{R}
\end{equation}
Taking everything together, we want to find $E$ that maximizes the mean fidelity
\begin{equation}
F=\frac{{\mathrm{Tr}}(ER)}{{\mathrm{Tr}}(EA)},
\label{Fmean}
\end{equation}
where $R\geq 0$ and $A >0$ are defined above. The positive semidefinite
operator $E$ representing a trace-decreasing CP
map must satisfy the constraint (\ref{tracedecreasing}). 
However, this constraint is irrelevant as far as the mean 
fidelity (\ref{Fmean}) is concerned because the value of $F$ does
not change under the re-normalization 
\begin{equation}
E\rightarrow E'=e_{\mathrm{max}}^{-1} E , 
\label{normalization}
\end{equation}
where $e_{\mathrm{max}}=
{\mathrm{max}}[{\mathrm{eig}}({\mathrm{Tr}}_{\mathrm{out}}E)]$ 
and $E'$ satisfies the inequality (\ref{tracedecreasing}) by construction. 
This fact greatly simplifies the analysis. 
Strictly speaking, these arguments are valid only for
finite dimensional Hilbert spaces where
$e_{\mathrm{max}}$ is always finite and $\bar{P}'={\mathrm{Tr}}[E'A]>0$ since
$A>0$. As we will see in the next section, a little extra care is needed 
when dealing with infinite dimensional systems.

The above argumentation shows that we have to maximize the fidelity (\ref{Fmean})
under the constraint $E\geq 0$. Without loss of generality, we can
assume that the optimal $E$ is a pure state $E=|E\rangle\langle E|$.
We introduce new state $|\tilde{E}\rangle=A^{1/2}|E\rangle$ and rewrite
(\ref{Fmean}) as follows,
\begin{equation}
F=\frac{\langle \tilde{E}|A^{-1/2} RA^{-1/2}|\tilde{E}\rangle}%
{\langle\tilde{E}|\tilde{E}\rangle} .
\end{equation}
It follows that the optimal vector $|\tilde{E}\rangle$ is the
eigenvector $|\mu_{\mathrm{max}}\rangle$ of $M=A^{-1/2}RA^{-1/2}$ that 
corresponds to the maximal eigenvalue $\mu_{\mathrm{max}}$ of $M$.
The maximal achievable mean fidelity is
 equal to the maximal eigenvalue,
\begin{equation}
F_{\mathrm{max}}={\mathrm{max}}[{\mathrm{eig}}(M)] 
\equiv {\mathrm{max}}[{\mathrm{eig}}(A^{-1}R)].
\label{Fmax}
\end{equation}
This formula is one of the the main results of the present paper.  The 
transformation  that achieves $F_{\mathrm{max}}$ is explicitely given by
\begin{equation}
E=e_{\mathrm{max}}^{-1}
A^{-1/2}|\mu_{\mathrm{max}}\rangle\langle \mu_{\mathrm{max}}|A^{-1/2},
\end{equation}
where we have normalized according to (\ref{normalization}) so that $E$
is a trace-decreasing map. If the largest eigenvalue $\mu_{\mathrm{max}}$ 
is non-degenerate, 
then this is the unique optimal $E$ and the problem is thus completely solved.
However, if the eigenvalue $\mu_{\mathrm{max}}$ is $n$-fold degenerate,
with $|\mu_{\mathrm{max},j}\rangle$, $j=1,\ldots,n$  being the $n$
eigenvectors, then there exist many
different transformations that saturate the fidelity bound (\ref{Fmax}).
It can be proved by direct substitution into Eq. (\ref{Fmean}) that any operator
\begin{equation}
E=\sum_{j,k=1}^n E_{jk}A^{-1/2}|\mu_{\mathrm{max},j}\rangle\langle
\mu_{\mathrm{max},k}|A^{-1/2},
\end{equation}
yields the maximal fidelity $F=\mu_{\mathrm{max}}$.
Let $\mathcal{K}$ be the Hilbert space spanned by the vectors
$A^{-1/2}|\mu_{\mathrm{max},j}\rangle$. Then $E$ can be any positive
semidefinite operator on $\mathcal{K}$ that satisfies (\ref{tracedecreasing}).
In this case, we would like to find the  map $E$ that maximizes the
average probability of success $\bar{P}$ while reaching the fidelity 
$F_{\mathrm{max}}$.  The optimization problem that has to be solved can 
be formulated as follows,
\begin{eqnarray}
&\mathrm{maximize} \quad \bar{P}={\mathrm{Tr}}[E A]
\quad \mathrm{under~the~constraints} & \nonumber \\
& E\geq 0, \quad  E\in B(\mathcal{K}), \quad 
{\mathrm{Tr}}_{\mathrm{out}}[E]\leq \openone,&
\nonumber \\
\label{SDP}
\end{eqnarray}
where $B(\mathcal{K})$ denotes the set of linear bounded operators on
$\mathcal{K}$.
This is an instance of the so-called semidefinite program (SDP) that can be
very efficiently solved  numerically and by means of the duality lemma
one can easily check that the global maximum was found \cite{Vandenberghe96}.
 In this context it is worth noting
that many optimization problems in quantum information theory can be
formulated as semidefinite programs. This includes several separability
criteria \cite{Doherty02,Woerdeman03}, 
calculation of distillable entanglement \cite{Rains01,Verstraete02}, determination 
of optimal POVM for discrimination of quantum states \cite{Jezek02,Eldar03}, 
derivation of optimal trace-preserving 
CP maps for cloning \cite{Audenaert02,Fiurasek02}, construction of local 
hidden variable theories  \cite{Terhal03}  etc.

Generally, the fidelity $F(\rho_{\mathrm{in}})$ will depend 
on $\rho_{\mathrm{in}}$. However, there is an important class of sets 
of input states $S_{\mathrm{in}}$ and
transformations (\ref{inout}) such that the optimal CP map is universal. By
universality we mean that the probability of success $P(\rho_{\mathrm{in}})$
as well as the fidelity $F(\rho_{\mathrm{in}})$ is independent of  
$\rho_{\mathrm{in}}$. This occurs whenever the set of the input and output 
states can be obtained as orbits of some group $G$. Consider a compact group 
$G$ with elements $g$. Let $U(g)$ and $V(g)$ denote unitary representations
of $G$ on ${\mathcal{H}}_{\mathrm{in}}$ and ${\mathcal{H}}_{\mathrm{out}}$, 
respectively. The unitary $U(g)$ generates the set of input states,
\begin{equation}
\rho_{\mathrm{in}}(g)= U(g) \rho_{\mathrm{in}}(g_0) U^\dagger(g),
\label{rhoinorbit}
\end{equation}
where $g_0$ is the identity element of the group and $U(g_0)=\openone$.
We also assume that the set of output states can be
obtained from $\psi_{\mathrm{out}}(g_0)$ as follows,
\begin{equation}
\psi_{\mathrm{out}}(g)= V(g) \psi_{\mathrm{out}}(g_0) V^{\dagger}(g)
\end{equation}
and $V(g_0)=\openone$.
The final assumption is that the  distribution of the  inputs coincides with
the invariant measure on the group $G$, $d\rho_{\mathrm{in}}=dg$. Under these
assumptions, it is possible to convert any optimal map $\mathcal{E}$ into a
universal map $\tilde{\mathcal{E}}$ that achieves the same fidelity as
$\mathcal{E}$ by the twirling operation. 
One first applies randomly a unitary $U(h)$ to the input and then
this is undone by applying $V^{-1}(h)$ to the output.
The composition of the twirling operation with the map $\mathcal{E}$
yields
\begin{equation}
\rho_{\mathrm{out}}(g)= \int_G V^\dagger(h) 
\mathrm{Tr}_{\mathrm{in}}[E \rho_{\mathrm{in}}^T(hg)\otimes 
\openone_{\mathrm{out}} ] V(h) dh
\end{equation}
and the  probability of success reads
\begin{equation}
P'[\rho_{\mathrm{in}}(g)]=\int_G \mathrm{Tr}[E \rho_{\mathrm{in}}^T(hg) 
\otimes\openone_{\mathrm{out}}] dh = \bar{P}.
\end{equation}
Here, we used the group composition law $U(h)U(g)=U(hg)$ to obtain
 $U(h)\rho_{\mathrm{in}}(g)U^\dagger(h)=\rho_{\mathrm{in}}(hg)$,
and the substitution $q=hg$, $dq=dh$. Similarly, we find that
\begin{equation}
F'(\rho_{\mathrm{in}})= \frac{1}{\bar{P}}\int_G
{\mathrm{Tr}}[E \rho_{\mathrm{in}}^T(hg)\otimes \psi_{\mathrm{out}}(hg)] dh= F,
\end{equation}
which confirms that the twirling operation results in a universal
machine that works equally well for all possible input states.

\section{Probabilistic cloning}

Having established the general formalism, we now turn our attention to
the explicit examples of application. Let us first consider the
universal symmetric $1\rightarrow M$ cloning machine for qubits.
Here, the input state is a single qubit
$|\psi\rangle=\cos\frac{\vartheta}{2}|0\rangle
+e^{i\phi}\sin\frac{\vartheta}{2}|1\rangle$,
uniformly distributed over the surface of the Bloch sphere,
$d\psi_{\mathrm{in}}=\frac{1}{4\pi} \sin\vartheta d\vartheta d\phi$. The cloning machine
should produce $M$ identical clones, hence 
$\psi_{\mathrm{out}}=\psi_{\mathrm{in}}^{\otimes M}$.  
The operators $R$ and $A$
can be easily calculated with the use of the Schur lemma,
\begin{equation}
R=\frac{1}{M+2}(\Pi_{+,M+1})^{T_{1}},
\qquad A=\frac{1}{2} \openone_1 \otimes \Pi_{+,M},
\label{RAuniversalclone}
\end{equation}
where $\Pi_{+,M}$ denotes a projector onto symmetric subspace of $M$ qubits
and $T_1$ indicates partial transposition with respect to the first qubit.
On inserting the operators (\ref{RAuniversalclone}) into Eq. (\ref{Fmax}) we find that 
$F_{\mathrm{max}}=2/(M+1)$. As shown in Ref. \cite{Werner98} the optimal
deterministic cloning machine saturates this bound, hence it is impossible 
to improve the fidelity via probabilistic cloning. 

Let us now consider the transposition operation for qudits, i.e. a map that
produces a transposed qudit state $\psi^T$ (in some fixed basis) 
from $N$ copies of $\psi$,  $\psi^{\otimes N}\rightarrow \psi^T$. 
For qubits, this map is unitarily equivalent to the 
universal NOT gate $\psi^{\otimes N}\rightarrow \psi_\perp$ 
\cite{Gisin99,Buzek99NOT}.
The Hilbert space ${\mathcal{H}}_{\mathrm{in}}$ is the fully symmetric subspace 
of the Hilbert space of $N$  qudits and ${\mathcal{H}}_{\mathrm{out}}$ 
is the Hilbert space of a single qudit.  In the formulas (\ref{lambda}) and
(\ref{R}) for $A$ and $R$ we average over all $\psi_{\mathrm{in}}$ that are
represented as orbits of the group $SU(d)$ according to Eq. (\ref{rhoinorbit}).
The probability density $d\psi_{\mathrm{in}}\equiv dg$, where $dg$ is the
invariant measure on the group $SU(d)$. With the use of the Schur lemma 
one easily finds
\begin{equation}
A= \frac{1}{D(N,d)} \openone, \qquad 
R=\frac{1}{D(N+1,d)} \Pi_{+,N+1}^{(d)},
\end{equation}
 where
$\Pi_{+,N+1}^{(d)}$ is the projector onto symmetric subspace of $N+1$ qudits 
and $D(N+1,d)={N+d \choose d-1}$ is the dimension of this subspace.
The optimal fidelity obtained from Eq. (\ref{Fmax}) reads
$F_{\mathrm{max}}=(N+1)/(N+d)$, which is exactly the fidelity of the optimal
{\em deterministic} estimation of the qudit state from $N$ copies
\cite{Bruss99}.  Note that the fidelity $F_{\mathrm{max}}=2/(d+1)$ of the 
optimal deterministic  transposition map for  $N=1$ was recently 
derived in \cite{Buscemi03}.

In all the above examples the optimal probabilistic machine could not
outperform the deterministic machines. This can be
attributed to the very high symmetry present in all the above considered
examples. The question is whether there are interesting cases when 
$S_{\mathrm{in}}$ is a continuous set of linearly dependent states and the
probabilistic machine achieves higher fidelity than the deterministic one. 
Below we answer this question in affirmative by providing explicit  examples.
We will focus on phase covariant cloning machines, where the underlying
group is the Abelian group $U(1)$. Specifically, we shall first consider
probabilistic $N\rightarrow M$ phase covariant cloning of qubits
\cite{Bruss00,DAriano03}. Here, the
input state $|\psi\rangle=(|0\rangle+e^{i\phi}|1\rangle)/\sqrt{2}$
lies on the equator of the Bloch sphere and is characterized by a
single parameter, the phase  $\phi$. Moreover, 
$\int d g=\int_0^{2\pi} d\phi/(2\pi)$. The input and output states are given by
\begin{equation}
\psi_{\mathrm{in}}=\psi^{\otimes N}, 
\qquad \psi_{\mathrm{out}}=\psi^{\otimes M}.
\end{equation}
The integrals appearing in the expression (\ref{R})
 can easily be carried out and one arrives at
\begin{equation}
R=\frac{1}{2^{M+N}}\sum_{y=-N}^{M} |\Phi_{M,N,y}\rangle \langle
\Phi_{M,N,y}|,
\end{equation}
where
\begin{equation}
|\Phi_{M,N,y}\rangle=\sum_{k=\max(0,-y)}^{\min(N,M-y)}
\sqrt{{N \choose k} {M \choose k+y}}
|N,k\rangle |M,y+k\rangle,
\end{equation}
and $|N,k\rangle$ denotes a totally symmetric state of $N$ qubits with
$k$ qubits in the state $|1\rangle$ and $N-k$ qubits in the state $|0\rangle$.
Similarly, one gets
\begin{equation}
A=\frac{1}{2^N}\sum_{k=0}^N {N \choose k} |N,k\rangle\langle N,k|\otimes
\openone_{\mathrm{out}}.
\end{equation}
Since the states $|\Phi_{M,N,y}\rangle$  are mutually orthogonal,
the matrix $R$ is diagonal and, consequently, also $M=A^{-1/2}RA^{-1/2}$
is diagonal. The maximal eigenvalue can thus be easily determined and
the maximal fidelity is given by
\begin{equation}
F_{\mathrm{max}}(N,M)= \frac{1}{2^M}\sum_{k=0}^N
{M \choose k+ \left[\frac{M-N}{2}\right]},
\label{Fphase}
\end{equation}
where $[x]$ denotes the integer part of $x$. For $N>1$, the fidelity
(\ref{Fphase}) is higher than the fidelity of the optimal 
deterministic phase covariant cloner that was given in \cite{DAriano03}. 
The improvement  of the fidelity is typically of the order of one percent.

The optimal probabilistic cloning transformation can be written as
\begin{equation}
|N,k\rangle \rightarrow \frac{1}{\cal{N}} 
\sqrt{{M \choose k+\Delta_{MN}}{N \choose k}^{-1}} \,
|M,\Delta_{MN}+k\rangle,
\label{cloningmap}
\end{equation}
where $\Delta_{MN}=\left[\frac{M-N}{2}\right]$ and
\begin{equation}
\mathcal{N}=\max_k \sqrt{{M \choose k+\Delta_{MN}} {N \choose k}^{-1} }
\label{Nfactor}
\end{equation}
is a normalization prefactor. If $M-N$ is even, then Eq. (\ref{cloningmap})
is the unique optimal phase-covariant probabilistic cloning transformation
that optimally matches the input state $|\psi\rangle^{\otimes N}$ onto the 
ideal  output $|\psi\rangle^{\otimes M}$. For odd $M-N$, however, 
we can obtain another optimal operation by replacing $\Delta_{MN}$ with
$\Delta_{MN}+1$ in Eqs. (\ref{cloningmap}) and (\ref{Nfactor}). This implies 
that the support $\mathcal{K}$ 
of the optimal operator $E$ is two dimensional and the optimal map that
maximizes the success probability has to be calculated by solving the
semidefinite program (\ref{SDP}). 

A second example where the probabilistic cloner outperforms the deterministic
one is the $1\rightarrow M$ copying of coherent 
states $|\alpha\rangle$ on a circle. Recall that
$|\alpha\rangle=e^{-|\alpha|^2/2}\sum_{n=0}^\infty \alpha^N/\sqrt{n!}\,|n\rangle$, 
where $|n\rangle$ is the $n$-photon Fock state.  Since we assume 
that $|\alpha\rangle$ lie on a circle, the amplitude $r=|\alpha|$ is  
fixed while the phase $\phi=\arg(\alpha)$ is arbitrary. First of all, we observe that the 
perfect cloning is equivalent to noiseless amplification, because the output
state $|\alpha\rangle^{\otimes M}$ can be unitarily mapped onto the state 
$|\sqrt{M}\alpha\rangle\otimes |0\rangle^{\otimes M-1}$ by an array of $M-1$
beam splitters \cite{Fiurasek01CV}. Thus the cloning  is equivalent to
$|\alpha\rangle\rightarrow |\sqrt{M}\alpha\rangle$. The operators $A$
and $R$ are calculated as averages over the phase $\phi$ and one finds that the
maximum eigenvalue of $M$ is $\mu_{\mathrm{max}}=1$ which indicates that an exact
probabilistic cloning is possible. However, we must be
careful because we deal with infinite dimensional Hilbert space and it turns out
the the fidelity $F= 1$ can be achieved only in the limit of zero
probability of success $P\rightarrow 0$. Nevertheless, arbitrarily high fidelity
can be reached with finite success probability if we first project the input 
state onto the subspace spanned by the first $N+1$ Fock states
$|0\rangle,\ldots,|N\rangle$ and then apply a diagonal filter that approximates
the noiseless amplification, $|n\rangle \rightarrow M^{(n-N)/2}|n\rangle$,
$n=0,\cdots,N$. The fidelity 
\begin{equation}
F=e^{-M |\alpha|^2}\sum_{n=0}^N \frac{M^n|\alpha|^{2n}}{n!},
\end{equation}
can be arbitrarily close to $1$ as  $N\rightarrow \infty$. Clearly, this
probabilistic cloning  achieves even higher fidelity for the coherent states
inside the circle, where $|\alpha|<r$.

\section{Purification of mixed states}

Another important application of the optimization technique developed 
in Sec. II consists  in the design of the optimal protocols  for purification 
of mixed quantum states.  Suppose that Alice and Bob can communicate
via a noisy quantum channel $\mathcal{C}$.
Alice wants to send to Bob a quantum state $\psi$ from some set 
$S_{\mathrm{in}}$.  However, since the channel is noisy,  Bob receives mixed state 
$\rho={\mathcal{C}}(\psi)$. To partially compensate for the effects of the noisy 
channel, Alice sends $N$ copies of the state $\psi$ to Bob who subsequently 
attempts to extract $\psi$ from the state $\rho^{\otimes N}$.
The purification of qubits transmitted through the
depolarizing channel, $\rho= \eta\psi +\frac{1}{2}(1-\eta) \openone$, 
has been analyzed in detail in Refs. \cite{Cirac99,Keyl01}. Very recently, 
the optimal purification protocol for two copies of the qubit has 
been demonstrated experimentally for the polarization states of single photons 
by exploiting the interference of two photons on a balanced beam splitter \cite{Ricci04}.  
Applications  of the purification procedure to the quantum state estimation 
and transmission have been discussed in Refs. \cite{Mack01,Fischer01}.
Here, we demonstrate that the present optimization procedure can be used to
straightforwardly determine the optimal probabilistic purification protocol.
Then we will consider the same problem for the amplitude damping channel.

\subsection{Depolarizing channel}

For the sake of simplicity we illustrate the method on the case when Alice
sends two copies of the state $|\psi\rangle$ to Bob, hence  Bob's input
mixed state reads 
$\rho_{\mathrm{in}}=(\eta \psi+\frac{1}{2}(1-\eta)\openone)^{\otimes2}$.
The ideal output is a single-qubit pure state $\psi$. Assuming uniform
distribution of $\psi$ over the surface of the Bloch sphere, one obtains the
following expressions for the operators $A$ and $R$,
\begin{eqnarray}
A&=&\left(\frac{1}{3}\eta^2
\Pi_{+,12}+\frac{1-\eta^2}{4}\openone_1\otimes\openone_2\right)\otimes \openone_3,
\nonumber \\
R^{T_3}&=&\frac{\eta^2}{4}\Pi_{+,123}+\frac{(1-\eta)^2}{8} \openone_{123}
\nonumber \\ & & 
+\frac{\eta}{6}(1-\eta)(\openone_1\otimes\Pi_{+,23}
+\openone_2\otimes \Pi_{+,13}),
\end{eqnarray}
where $1$ and $2$ label the input qubits while $3$ labels the output qubit, 
$\Pi_{+,jk}$ and $\Pi_{+,ijk}$ are projectors on the symmetric subspace of two
qubits $i,j$ or three qubits $i,j,k$, respectively, 
and $T_3$ stands for the partial transposition with respect to the third qubit.
From Eq. (\ref{Fmax}) we obtain the maximal achievable purification fidelity:
\begin{equation}
F=\frac{3+4\eta+\eta^2}{2(3+\eta^2)},
\label{Fdepolarizing}
\end{equation}
which is larger than the original fidelity  
$F_0=\langle\psi|\rho|\psi\rangle=(1+\eta)/2$,  for all $0<\eta<1$. 
As shown in Refs. \cite{Cirac99,Keyl01,Ricci04}, the optimal purification strategy 
is to project the two-qubit state $\rho^{\otimes 2}$ onto the symmetric 
subspace and then throw away one of the qubits. This procedure achieves the
optimal fidelity (\ref{Fdepolarizing}). Let us now demonstrate that this protocol
can be derived by solving the semidefinite program (\ref{SDP}). The maximal
eigenvalue of the matrix $M$ is doubly degenerate and the basis states that
span the two dimensional  Hilbert space $\mathcal{K}$ are given by
\begin{eqnarray}
|e_{1}\rangle=\frac{1}{\sqrt{6}}(\sqrt{2}|\Psi_+\rangle|0\rangle+2|11\rangle|1\rangle),
\nonumber \\
|e_{2}\rangle=\frac{1}{\sqrt{6}}(\sqrt{2}|\Psi_+\rangle|1\rangle+2|00\rangle|0\rangle),
\end{eqnarray}
where $|\Psi_+\rangle=(|01\rangle+|10\rangle)/\sqrt{2}$.
From Eq. (\ref{P}) and the definition of $A$ we find that 
$\bar{P}={\mathrm{Tr}}_{\mathrm{in}}[\lambda X]$, where 
$\lambda=\int_{S_{\mathrm{in}}}\rho_{\mathrm{in}}^T d\rho_{\mathrm{in}}$ and
$X={\mathrm{Tr}}_{\mathrm{out}}(E)$. 
Since $E=\sum_{j=1,2}c_{jk}|e_j\rangle\langle e_{k}|$ it follows that
the support of $X$ is the symmetric subspace of two
qubits. From Eq. (\ref{tracedecreasing}) we thus have $X\leq \Pi_{+,12}$.
The maximum $\bar{P}$ is obtained when $X=\Pi_{+,12}$, which can be achieved 
by the following choice of $E$:
\[
E=\frac{3}{2}(|e_1\rangle\langle e_1|+|e_2\rangle\langle e_2|).
\]
It is easy to check that this trace-decreasing CP map indeed describes the
projection of two qubits onto the symmetric subspace followed by tracing over
the second qubit.

\begin{figure}
\center{\psfig{figure=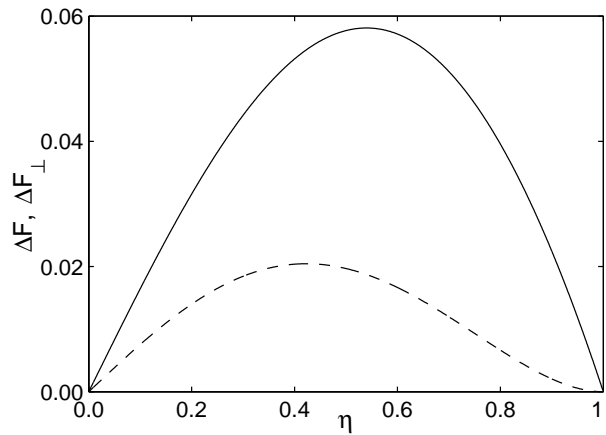,width=0.95\linewidth}}
\caption{The maximal possible improvements of fidelity $\Delta F(\eta)$ (solid line) and 
$\Delta F_{\perp}(\eta)$ (dashed line) that can be achieved by purification 
when the  two-qubit state $|\psi\psi\rangle$ or $|\psi\psi_\perp\rangle$, respectively,  
is sent through the depolarizing channel with parameter $\eta$.}
\end{figure}

The strategy to send the state $|\psi\psi\rangle$ is not the only
possible option how Alice can encode the state $|\psi\rangle$ into the two
qubits that she sends to Bob  via depolarizing channel. 
For instance,  she can send him the state $|\psi\rangle|\psi_{\perp}\rangle$,
where $\langle \psi|\psi_\perp \rangle=0$.
As shown by Gisin and Popescu \cite{Gisin99}, the state $|\psi\rangle$ can 
be estimated with higher fidelity from a single copy of the state 
$|\psi\psi_\perp\rangle$ than from a single copy of  $|\psi\psi\rangle$. 
We cannot therefore a-priori rule out that sending the state 
$|\psi\psi_{\perp}\rangle$ can be advantageous also in the present context. 
If Alice sends $|\psi\psi_\perp\rangle$ then Bob's input mixed state reads
\[
\rho_{\mathrm{in}}=
\left(\eta\psi+\frac{1-\eta}{2}\openone\right)\otimes
\left(\eta\psi_{\perp}+\frac{1-\eta}{2} \openone\right).
\]
The calculation of the optimal purification fidelity is completely similar to
the case of sending $|\psi\psi\rangle$. The integrals (\ref{lambda}) and
(\ref{R})  yielding the relevant  operators $A_{\perp}$ and $R_{\perp}$ can 
be easily  evaluated with the help of the substitution 
$\psi_{\perp}=\openone - \psi$  and we obtain
\begin{eqnarray}
A_{\perp}&=&\left(\frac{1+\eta^2}{4}\openone_1\otimes\openone_2
- \frac{1}{3}\eta^2 \Pi_{+,12} \right)\otimes \openone_3,
\nonumber \\
R_{\perp}^{T_3}&=&-\frac{\eta^2}{4}\Pi_{+,123}+\frac{1-\eta^2}{8} \openone_{123}
\nonumber \\ & & 
+\frac{\eta}{6}(1+\eta)\openone_2\otimes\Pi_{+,13}
-\frac{\eta}{6}(1-\eta)\openone_1\otimes \Pi_{+,23},
\nonumber \\
\end{eqnarray}
The maximal fidelity $F_{\perp}$ is determined as the maximum eigenvalue of 
the operator $A_{\perp}^{-1}R_{\perp}$.

The results of numerical calculations
are given in Fig. 1. For comparison we plot on this figure the gains 
in fidelity $\Delta F_{\perp}=F_{\perp}(\eta)-F_{0}(\eta)$ and 
$\Delta F=F(\eta)-F_{0}(\eta)$  achieved when Alice sends the state 
$|\psi\psi_\perp\rangle$ or $|\psi\psi\rangle$, respectively. 
We can see that as far as the
purification is concerned, it is strictly better for any $0<\eta <1$ 
to send the state $|\psi\psi\rangle$ than $|\psi\psi_{\perp}\rangle$. 
The non-zero values of $\Delta F_{\perp}(\eta)$
clearly show that purification is possible also when Alice sends
the state $|\psi\psi_{\perp}\rangle$ but the fidelity improvement is 
much smaller than when sending the state $|\psi\psi\rangle$.

\subsection{Amplitude damping channel}

To further illustrate the utility and universality of our optimization method,
let us now consider a different class of noisy channels, namely, an amplitude-damping 
channel that maps a pure state $\psi$ onto a mixed state
\begin{equation}
\rho_{AD}(\vartheta,\phi)=
\left(
\begin{array}{cc}
\eta^2\cos^2\frac{\vartheta}{2} & \frac{\eta}{2}\sin\vartheta\, e^{-i\phi} \\[2mm]
\frac{\eta}{2}\sin\vartheta \, e^{i\phi}  & 1-\eta^2\cos^2\frac{\vartheta}{2}
\end{array}
\right).
\end{equation}
This channel may arise, for instance,  when the qubit is represented by the 
ground and excited atomic states $|g\rangle$ and $|e\rangle$ where $|e\rangle$
can decay to $|g\rangle$ via spontaneous emission. 
In order to preserve the covariance (\ref{rhoinorbit}) that guarantees 
the universality of  the optimal purification protocol, 
we shall assume that  Alice is sending to Bob 
$N$ copies of a state
$|\psi(\phi)\rangle=(|0\rangle+e^{i\phi}|1\rangle)/\sqrt{2}$ that 
lies on the equator of the Bloch sphere, i.e. the set $S_{\mathrm{in}}$ 
consists of the states  
$\rho_{BOB}^{\otimes N}(\phi)=\rho_{AD}^{\otimes N}(\pi/2,\phi)$.

The operators $R_{AD}$ and $A_{AD}$ are obtained by integrating over the phase
$\phi$,
\begin{eqnarray}
A_{AD}&=&\frac{1}{2\pi}\int_{0}^{2\pi} \rho_{BOB}^{\otimes N}(-\phi) 
\otimes \openone_{\mathrm{out}} \, d\phi,
\nonumber \\
R_{AD}&=&\frac{1}{2\pi}\int_{0}^{2\pi} \rho_{BOB}^{\otimes N}(-\phi) 
\otimes \psi(\phi) \, d\phi.
\end{eqnarray}
We now prove that when determining the maximum achievable purification fidelity, 
we can assume that the optimal map $\mathcal{E}$ is a composition of two maps, 
$\mathcal{E}=\tilde{\mathcal{E}}\circ\mathcal{P}_{N}$, where the map
$\mathcal{P}_N$ projects the input state $\rho_{BOB}^{\otimes N}$ onto the
symmetric subspace of $N$ qubits $\mathcal{H}_{+,N}$ and $\tilde{\mathcal{E}}$
maps operators on $\mathcal{H}_{+,N}$ onto operators on the Hilbert space of
single output qubit.
By definition, the operators $A_{AD}$ and $R_{AD}$ commute with
arbitrary  permutation  operator $\Pi_{j}$ that changes the order of the $N$
input qubits. Suppose that $E=|e\rangle\langle e|$ is an optimal map yielding
maximal fidelity $F_{\mathrm{max}}$ which is equal to the maximum eigenvalue 
of matrix  $M_{AD}=A_{AD}^{-1/2}R_{AD} A_{AD}^{-1/2}$. As shown in Sec. II, 
the corresponding optimal eigenvector $|\mu\rangle$ of $M$ is related to 
$|e\rangle$ as follows,  $|\mu\rangle=A_{AD}^{1/2}|e\rangle$. 
Let us now consider a  symmetrized state $|x\rangle$ 
that we obtain from $|e\rangle$ by making a linear superposition of 
all $N!$ permutations of $N$ input qubits,
\begin{equation}
|x\rangle=\sum_{j} \Pi_{j}\otimes\openone_{\mathrm{out}} |e\rangle.
\end{equation}
It is easy to show that $A_{AD}^{1/2}|x\rangle$ is an eigenstate of $M_{AD}$
with eigenvalue $F_{\mathrm{max}}$. We have
\begin{eqnarray}
M_{AD} A_{AD}^{1/2}\,|x\rangle&=&A_{AD}^{1/2}A_{AD}^{-1}R_{AD} \sum_{j} \Pi_j
A_{AD}^{-1/2}|\mu\rangle \nonumber \\
&=& A_{AD}^{1/2}\sum_{j} \Pi_j A_{AD}^{-1/2} F_{\mathrm{max}}|\mu\rangle
\nonumber \\ 
&=&F_{\mathrm{max}} A_{AD}^{1/2}|x\rangle, 
\end{eqnarray}
where we have used that both $A_{AD}$ and $R_{AD}$ commute with $\Pi_j$. 
Thus the map $X=|x\rangle\langle x|$ achieves the maximal fidelity
$F_{\mathrm{max}}$. It holds that 
$\mathrm{Tr}_{\mathrm{out}}(X) \in B(\mathcal{H}_{+,N})$ which  proves that we can restrict our attention to the maps
$\tilde{\mathcal{E}}$ when calculating the maximal fidelity of purified state.

The calculations can be further considerably simplified by the observation 
that the optimal map $\tilde{\mathcal{E}}$ can be made phase-covariant 
\cite{DAriano01,DAriano03},  that is, invariant under the twirling operation,
\begin{equation}
\tilde{E}=\int_{0}^{2\pi} \frac{d\phi}{2 \pi} U^{*}(\phi) \otimes V({\phi}) 
\tilde{E} U^T(\phi) \otimes V^\dagger(\phi), 
\label{Etwirling}
\end{equation}
where $U(\phi)|N,k\rangle=e^{ik\phi}|N,k\rangle$, $k=0,\ldots,N$ and
$V|j\rangle=e^{ij\phi}|j\rangle$, $j=0,1$. This implies that the operator
$\tilde{E}$ can be expressed as a direct sum,
\begin{equation}
\tilde{E}=\bigoplus_{k=-1}^{N} \tilde{E}_{k},
\label{EdecompositionAD}
\end{equation}
where the support of the operator $\tilde{E}_{k}$ is a two-dimensional Hilbert space 
$\mathcal{H}_k$, $k=0,\ldots,N-1$  spanned by $|N,k\rangle|0\rangle$ and 
$|N,k+1\rangle|1\rangle$ and
\begin{eqnarray}
\tilde{E}_{-1}&=& \mu_{-1} |N,0\rangle\langle N,0| \otimes 
|1\rangle\langle 1|,
\nonumber \\
\tilde{E}_{N+1}&=&\mu_{N+1} |N,N\rangle\langle N,N| \otimes  
|0\rangle\langle 0|.
\end{eqnarray}
The decomposition (\ref{EdecompositionAD}) implies that we can perform the optimization of each CP
map $\tilde{E}_k$  separately and then choose the $k$ that yields the highest fidelity.
It is easy to see that the trace decreasing CP maps $\tilde{E}_{-1}$ and
$\tilde{E}_{N+1}$
lead to very low purification fidelity $\frac{1}{2}$ so it is optimal for all $N$
to set $\mu_{-1}=\mu_{N+1}=0$. Without loss of generality, we can assume that
the optimal $\tilde{E}_{k}$ is a rank-one operator, 
$\tilde{E}_{k}=|\tilde{E}_k\rangle\langle \tilde{E}_k|$, where
\begin{equation}
|\tilde{E}_k\rangle=|N,k\rangle|0\rangle+\alpha_{N,k}(\eta)|N,k+1\rangle|1\rangle.
\label{Ekmap}
\end{equation}
The action of this operation can be understood as follows. First the $N$-qubit 
state is projected onto two-dimensional subspace of $\mathcal{H}_{+,N}$
spanned by $|N,k\rangle$  and $|N,k+1\rangle$ and then the following 
transformation is carried out, 
\begin{equation}
|N,k\rangle \rightarrow|0\rangle,  \qquad
|N,k+1\rangle \rightarrow
\alpha_{N,k}(\eta)|1\rangle.
\label{Nkqubit}
\end{equation}
The un-normalized density matrix of the purified qubit obtained 
by applying the map (\ref{Nkqubit})  reads
\begin{equation}
\rho_{\mathrm{out}}=
\left(
\begin{array}{cc}
\sigma_{k,k}^{(N)} & \alpha_{N,k}^{\ast} \sigma_{k,k+1}^{(N)} e^{-i\phi} \\[2mm]
\alpha_{N,k} \sigma_{k,k+1}^{(N)} e^{i\phi} & |\alpha_{N,k}|^2
\sigma_{k+1,k+1}^{(N)}
\end{array}
\right).
\label{rhopurifiedAD}
\end{equation}
The relevant matrix elements 
\[
\sigma_{j,k}^{(N)}=\langle N,j|\rho_{AD}^{\otimes N}(\pi/2,0) |N,k\rangle
\]
 can be expressed in terms of a finite series,
\begin{eqnarray*}
\sigma_{k,k}^{(N)}&=&  2^{-N}\eta^{2N-2k}  \sum_{l=0}^k
{ k \choose l} { N-k \choose k-l} (2-\eta^2)^l , \\
\sigma_{k,k+1}^{(N)} &=& 2^{-N} \eta^{2N-2k-1} \sqrt{\frac{k+1}{N-k}}
\nonumber \\ 
& & \times \sum_{l=0}^k {k \choose l} { N-k \choose k+1-l} (2-\eta^2)^l.
\end{eqnarray*}

The optimal $\alpha_{N,k}(\eta)$
that maximizes the fidelity of the purified state (\ref{rhopurifiedAD}) 
with respect to the original pure state 
$(|0\rangle+e^{i\phi}|1\rangle)/\sqrt{2}$ is given by
$\alpha_{N,k}=\sqrt{\sigma_{k,k}^{(N)}/\sigma_{k+1,k+1}^{(N)}}$
and the fidelity of purified qubit reads
\begin{equation}
F_{N,k}=\frac{1}{2}\left(1+\frac{\sigma_{k,k+1}^{(N)}}%
{\sqrt{\sigma_{k,k}^{(N)}\sigma_{k+1,k+1}^{(N)}}}\right).
\label{Fsigma}
\end{equation}
The maximum achievable fidelity can be found as a maximum over all $k$,
$F_{N,\mathrm{max}}=\max_{k}F_{N,k}$. Based on numerical calculations we
conjecture that for odd $N$ the best fidelity is
reached for $k=(N-1)/2$ while for even $N$ there are two alternatives 
leading to the same optimal $F$, namely $k=N/2-1$ and $k=N/2$.
For $N\leq 10$ we have checked that these choices of $k$ are optimal 
which supports this conjecture.

We shall now  present explicit results for $N=1,2,3$. Besides of maximal
fidelity, we are interested also in the maximal probability $\bar{P}$ 
of optimal purification. We have therefore carried out full calculations 
of the operators $A_{AD}$ and $R_{AD}$ for $N=1,2,3$ and determined the
degeneracy of the maximal eigenvalue of matrix $M$. These calculations reveal
that the optimal eigenvectors $A_{AD}^{1/2}|e\rangle$ of $M$ all satisfy the relation
$\mathrm{Tr}_{\mathrm{out}}[e]\in B(\mathcal{H}_{+,N})$ 
so we can in fact consider only the maps of the form (\ref{Ekmap}) 
without any loss of generality.

If only a single qubit is sent to Bob, then the only option is $k=0$ and 
the best filter obtained by setting  $\alpha_{1,0}=\eta/\sqrt{2-\eta^2}$
which achieves a fidelity $F_1=\frac{1}{2}(1+1/\sqrt{2-\eta^2})$.
The purification succeeds with probability $P_{1}=\eta^2$.

For $N=2$ the maximum fidelity  $F_2$ is given by 
\begin{equation}
F_2= \frac{1}{2}\left(1+\sqrt{\frac{2}{3-\eta^2}}\right).
\end{equation}
One option how to achieve $F_2$ is to choose $k=0$ 
and $\alpha_{2,0}=\eta/\sqrt{3-\eta^2}$. The second alternative is $k=1$
and $\alpha_{2,1}=\eta\sqrt{3-\eta^2}/(2-\eta^2)$. The Hilbert space $\mathcal{K}$
of the admissible optimal operations $E$ is thus two-dimensional and spanned by
basis states 
\begin{eqnarray}
|e_1\rangle=|00\rangle|0\rangle+\alpha_{2,0}|\Psi_+\rangle|1\rangle,
\nonumber \\
|e_2\rangle=|\Psi_+\rangle|0\rangle+\alpha_{2,1}|11\rangle|1\rangle.
\end{eqnarray}
To find $E$ that maximizes $\bar{P}$ we must solve (\ref{SDP}).

It follows from (\ref{Etwirling}) and (\ref{EdecompositionAD})
that the optimal $E$ is diagonal in basis $|e_1\rangle,|e_2\rangle$,
$E=p_1 e_1+ p_2 e_2$. Consequently, the semidefinite program (\ref{SDP}) 
reduces  to a linear program and we have to maximize
\[
\bar{P}=\frac{1}{2} p_1 \eta^4+\frac{1}{2} p_2\eta^2(3-\eta^2)
\]
under the constraints
\[
0\leq p_1\leq 1, \qquad
0\leq p_2 \leq \alpha_{2,1}^{-2} , \qquad
0\leq p_1\alpha_{2,0}^2+p_2\leq 1.
\]
For $\eta\leq \eta_{\mathrm{th}}\equiv (7-\sqrt{17})^{1/2}/2$ 
the optimal coefficients read $p_1=0$, $p_2=1$  while 
for $\eta>\eta_{\mathrm{th}}$
we have $p_1=(\alpha_{2,1}^2-1)/(\alpha_{2,0}^2\alpha_{2,1}^2)$ and
$p_2=\alpha_{2,1}^{-2}$. For all $0<\eta\leq 1$ the optimal probability 
is given by a simple formula 
\begin{equation}
P_{2}=\frac{1}{2}\eta^2(3-\eta^2).
\end{equation}

\begin{figure}[t]

\centerline{\psfig{figure=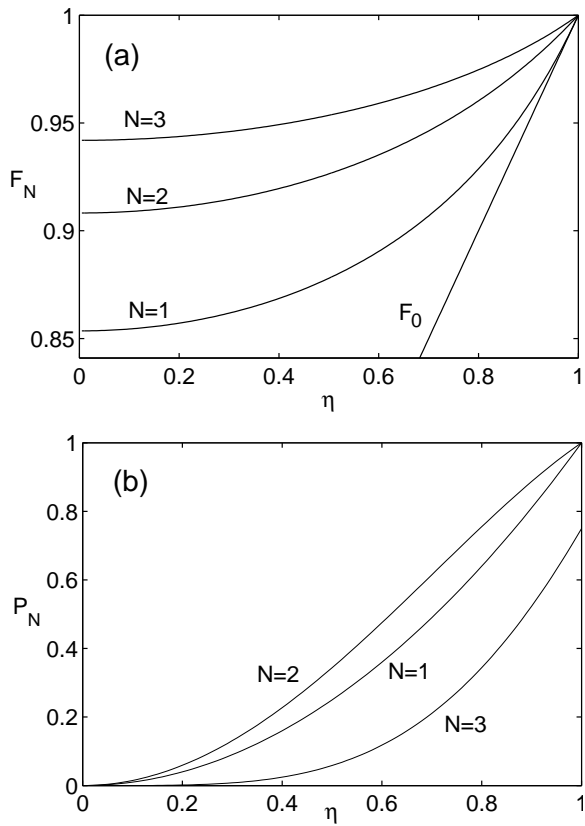,width=0.9\linewidth}}

\caption{ (a) Maximal fidelity of the purified qubit when Alice sends $N$ copies of
the qubit via amplitude damping channel parametrized by $\eta$. For comparison, 
the curve labeled $F_0$ displays the fidelity of the single qubit after 
passing through the channel, $F_0=(1+\eta)/2.$ (b) The corresponding maximal
probability of successful purification.}
\end{figure}

Finally, when Alice sends three qubits to Bob ($N=3$), then the 
optimal fidelity of Bob's purified qubit is given by
\begin{equation}
F_3=\frac{1}{2}\left[1+\frac{5-2\eta^2}{(4-\eta^2)\sqrt{2-\eta^2}}\right],
\end{equation}
and the only way to reach $F_3$ is to choose $k=1$ and 
$\alpha_{3,1}=\eta/\sqrt{2-\eta^2}$. The purification succeeds with probability 
$P_3=\eta^4-\eta^6/4.$

The dependence of the optimal fidelities on $\eta$ is plotted in Fig. 2(a)
which clearly illustrates that the purification results in a significant
improvement of the fidelity. The relative improvement is maximal when 
$\eta\rightarrow 0$ but this is reached at the expense of very low probability
of success, see Fig. 2(b). Note also that $\lim_{\eta\rightarrow
1}P_3=3/4$. If Bob possesses three noisy qubits and tries to extract one qubit,
then his optimal probabilistic strategy will have a finite probability of 
failure for arbitrarily low damping.

\section{Conclusions}

In this paper we have investigated the optimal {\em probabilistic}
realizations of several important quantum-information-processing tasks
such as the optimal cloning of quantum states and purification of mixed 
quantum states. We have derived a simple formula for the 
maximum achievable average fidelity and we have provided an explicit
prescription how to construct a trace-decreasing CP map that 
reaches the fidelity $F_{\mathrm{max}}$. We have demonstrated that the 
fidelity of  probabilistic cloning can be strictly higher than the maximal 
fidelity of deterministic cloning even if the set of the cloned states 
is linearly dependent and continuous. However, it should be stressed that 
this improvement in fidelity is achieved at the expense of a
certain fraction of unsuccessful events when the probabilistic transformation
fails and does not produce any output state. 

The optimal probabilistic maps may find a variety of applications. 
For instance, the phase covariant cloning is an efficient attack on 
several  quantum key distribution protocols. In particular, the 
$2\rightarrow 3$ phase-covariant cloning is explored for eavesdropping 
purposes  in Ref. \cite{Acin03}. Thus, the probabilistic
phase-covariant cloning discussed in the present paper may be possibly 
used as a new eavesdropping attack. Moreover, the general theory of optimal 
probabilistic transformations developed in the
present paper has much broader range of applications than just cloning. 
In particular, it provides a method to engineer optimal protocols for
purification of mixed quantum states. 

We have seen on the example of the amplitude damping channel that the 
optimal probabilistic purification may result in a dramatic improvement 
of the fidelity of the final Bob's state with respect to the original 
state that was sent to him by Alice via a noisy channel. However, 
the large improvement of the fidelity is typically accompanied by a very 
low probability of success. It is therefore highly desirable to optimize the 
probabilistic transformation also with respect to the average success 
probability which leads to a semidefinite program that can be very
efficiently solved numerically. For the particular cases of purification 
of mixed states investigated in the present paper, we have been able 
to solve the resulting SDP analytically, by exploiting the symmetries inherent 
to the problem.

The protocol considered in the present paper can be even further generalized
as follows.  One can imagine a scenario where the average fidelity of the 
operation $F$ is maximized for a fixed chosen
average probability of success $\bar{P}$, or vice versa, these two
alternatives are clearly equivalent. Generally, there will always be 
a trade-off between $\bar{P}$ and $F$  and the optimal
fidelity will be some function of $\bar{P}$.
One can then choose the working point on the $F(\bar{P})$ curve that 
is most fitting for the particular task at hand. 
The determination of maximal $F$ obtainable for some fixed $\bar{P}$ 
can be formulated as a semidefinite program similar to that given 
by Eq. (\ref{SDP}).  The deterministic machines and the probabilistic 
machines that achieve the maximum possible fidelity represent 
two extreme regimes of such a more general scenario.

\vspace*{-5mm}

\acknowledgments

I would like to thank Nicolas J. Cerf for many stimulating discussions.
I acknowledge financial support from the Communaut\'e Fran\c{c}aise de
Belgique under grant ARC 00/05-251, from the IUAP programme of the Belgian
government under grant V-18, from the EU under project CHIC (IST-2001-33578) 
and  from the  grant LN00A015  of the Czech Ministry of Education.


\begin{thebibliography}{99}

\bibitem{Gisin02}
N. Gisin, G. Ribordy, W. Tittel, and H. Zbinden,
Rev. Mod. Phys. \textbf{74}, 145 (2002).


\bibitem{Shor94}
P. W. Shor, 
Proceedings of the 35th Annual Symposium on the Foundations 
of Computer Science,
p. 124, Ed. S. Goldwasser, IEEE Computer Society Press, New York, (1994).


\bibitem{Wootters82}
W.K. Wootters and W.H. Zurek, Nature (London) \textbf{299}, 802 (1982);
D. Dieks, Phys. Lett. \textbf{92A}, 271 (1982).


\bibitem{Buzek96}
V. Bu\v{z}ek and M. Hillery, Phys. Rev. A \textbf{54}, 1844 (1996).

\bibitem{Gisin97}
N. Gisin and S. Massar, Phys. Rev. Lett. \textbf{79}, 2153 (1997).

\bibitem{Bruss98}
D. Bruss, A. Ekert, and C. Macchiavello, 
Phys. Rev. Lett. \textbf{81}, 2598 (1998).



\bibitem{Werner98}
R.F. Werner, Phys. Rev. A \textbf{58}, 1827 (1998).

\bibitem{Buzek98}
V. Bu\v{z}ek  and M. Hillery, Phys. Rev. Lett. \textbf{81}, 5003 (1998).

\bibitem{Keyl99}
M. Keyl and R.F. Werner, J. Math. Phys. \textbf{ 40}, 3283 (1999);

\bibitem{Cerf01}
N.J. Cerf, J. Mod. Opt. \textbf{47}, 187 (2000).

\bibitem{Braunstein01}
S.L. Braunstein, V. Bu\v{z}ek, and M. Hillery,
Phys. Rev. A \textbf{ 63}, 052313 (2001);


\bibitem{Cerf00}
N.J. Cerf, A. Ipe, and X. Rottenberg, Phys. Rev. Lett. \textbf{ 85},
1754 (2000).

\bibitem{Braunstein01CV}
S.L. Braunstein {\em et al.}, Phys. Rev. Lett. \textbf{ 86}, 4938 (2001).

\bibitem{Fiurasek01CV}
J. Fiur\'{a}\v{s}ek, Phys. Rev. Lett. \textbf{86}, 4942 (2001).



\bibitem{Bruss00}
D. Bruss {\em et al.}, Phys. Rev. A \textbf{ 62}, 012302 (2000);

\bibitem{DAriano01}
G.M. D'Ariano and P. Lo Presti, Phys. Rev. A \textbf{ 64}, 042308 (2001);

\bibitem{DAriano03}
G.M. D'Ariano and C. Macchiavello, Phys. Rev. A \textbf{67}, 042306 (2003). 


\bibitem{Fiurasek03}
J. Fiur\'{a}\v{s}ek, Phys. Rev. A \textbf{67}, 052314 (2003). 


\bibitem{Navez03}     
P. Navez and N.J. Cerf, Phys. Rev. A \textbf{68}, 032313 (2003).


\bibitem{Lamoureux03}
L.-P. Lamoureux, P. Navez, J. Fiur\'{a}\v{s}ek, and N.J. Cerf, 
quant-ph/0302173v2.


\bibitem{Fuchs97}
C.A. Fuchs {\em et al.}, Phys. Rev. A \textbf{ 56}, 1163 (1997).

\bibitem{Niu99}
C.-S. Niu and R.B. Griffiths, Phys. Rev. A \textbf{ 60}, 2764 (1999).

\bibitem{Cerf02}
N.J. Cerf {\em et al.}, Phys. Rev. Lett. \textbf{ 88}, 127902 (2002).
 

\bibitem{Acin03}
A. Ac\'{\i}n, N. Gisin, and V. Scarani,  Phys. Rev. A \textbf{69}, 012309 (2004).

 
\bibitem{Duan98}
L.M. Duan and L.C. Guo, Phys. Lett. A \textbf{243},  261 (1998);
Phys. Rev. Lett. \textbf{80}, 4999 (1998).

\bibitem{Chefles98}
A. Chefles and S.M. Barnett, J. Phys. A:Math. Gen. \textbf{31}, 10097 (1998);
Phys. Rev. A \textbf{60}, 136 (1999).


\bibitem{Gisin99}
N. Gisin and S. Popescu, Phys. Rev. Lett. \textbf{83}, 432 (1999).

 
\bibitem{Buzek99NOT}
V. Bu\v{z}ek, M. Hillery, and R.F. Werner,
Phys Rev. A \textbf{60}, R2626 (1999).


\bibitem{Cirac99}
J.I. Cirac, A.K. Ekert, and C. Macchiavello,
Phys. Rev. Lett. \textbf{82}, 4344 (1999).

\bibitem{Keyl01}
M. Keyl and R.F. Werner,
Annales Henri Poincare \textbf{2}, 1 (2001).

\bibitem{Ricci04}
M. Ricci, F. De Martini, N.J. Cerf, R. Filip, J. Fiur\'{a}\v{s}ek, 
and  C. Macchiavello,  quant-ph/0403118.

\bibitem{Assumption}
We assume that there is no correlation between the device that prepares the
input states $\rho_{\mathrm{in}}$ and the degrees of freedom that subsequently
interact with $\rho_{\mathrm{in}}$.


\bibitem{Jamiolkowski72}
A. Jamiolkowski, Rep. Math. Phys. \textbf{ 3}, 275 (1972).


\bibitem{Vandenberghe96}
L. Vandenberghe and S. Boyd, SIAM  Rev. \textbf{38}, 49 (1996).


\bibitem{Doherty02}
A.\,C. Doherty, P.\,A. Parrilo, and F.\,M. Spedalieri,
Phys. Rev. Lett. \textbf{88}, 187904 (2002).

\bibitem{Woerdeman03}
H. J. Woerdeman Phys. Rev. A \textbf{67}, 010303 (2003). 

\bibitem{Rains01}
E.\,M. Rains, IEEE Trans. Inform. Theory \textbf{47},  2921 (2001).


\bibitem{Verstraete02}
F. Verstraete and H. Verschelde, Phys. Rev. A \textbf{66}, 022307 (2002);
Phys. Rev. Lett. \textbf{90}, 097901 (2003).



\bibitem{Jezek02}
M.~Je\v{z}ek, J.~\v{R}eh\'{a}\v{c}ek, and J.~Fiur\'{a}\v{s}ek,
Phys. Rev. A \textbf{65}, 060301(R) (2002).

\bibitem{Eldar03}
Y.C. Eldar, A. Megretski, and G.C. Verghese, 
IEEE Trans. Inform. Theory {\bf 49}, 1007 (2003).


\bibitem{Audenaert02}
K. Audenaert and B. De Moor, Phys. Rev. A \textbf{65}, 030302(R) (2002).



\bibitem{Fiurasek02}
J. Fiur\'{a}\v{s}ek, S. Iblisdir, S. Massar, and N.J. Cerf,
Phys. Rev. A \textbf{65}, 040302 (2002).


\bibitem{Terhal03}
B.M. Terhal, A.C. Doherty, and D. Schwab,
Phys. Rev. Lett. \textbf{90}, 157903 (2003). 


\bibitem{Bruss99}
D, Bruss and C. Macchiavello, Phys. Lett. A \textbf{253}, 249 (1999).


\bibitem{Buscemi03}
F. Buscemi, G. M. D'Ariano, P. Perinotti, M. F. Sacchi,
Phys. Lett. A \textbf{314}, 374 (2003).
 



\bibitem{Mack01}
H. Mack, D.G. Fischer, and M. Freyberger,
Phys. Rev. A \textbf{62}, 042301 (2001)


\bibitem{Fischer01}
D.G. Fischer, H. Mack, and M. Freyberger,
Phys. Rev. A \textbf{63}, 042305 (2001)



\end{thebibliography}
\end{document}